\documentclass[a4paper,twocolumn,showpacs,superscriptaddress]{revtex4}

\usepackage[english]{babel}
\usepackage{graphicx}
\usepackage{xspace}
\usepackage{longtable}
\usepackage{amsmath,amssymb}

\DeclareMathAlphabet{\mathbfit}{OT1}{cmr}{bx}{it}

\begin{document}

\title{Superconductivity in epitaxial thin films of
Na$_{\mathbf x}$CoO$_{\mathbf
2}\mathbf{\cdot}\mathbf{y}\mathbfit{D}_{\mathbf 2}$O}

\author{Y.~Krockenberger}
\affiliation{Max-Planck-Institute for Solid State Research,
Heisenbergstr.~1, 70569 Stuttgart, Germany}
\affiliation{Darmstadt University of Technology, Petersenstr.~23, 64287 Darmstadt,
Germany}
\author{I.~Fritsch}
\affiliation{Max-Planck-Institute for Solid State Research,
Heisenbergstr.~1, 70569 Stuttgart, Germany}
\author{G.~Christiani}
\affiliation{Max-Planck-Institute for Solid State Research,
Heisenbergstr.~1, 70569 Stuttgart, Germany}
\author{H.-U.~Habermeier}
\affiliation{Max-Planck-Institute for Solid State Research,
Heisenbergstr.~1, 70569 Stuttgart, Germany}
\author{Li Yu}
\affiliation{Max-Planck-Institute for Solid State Research,
Heisenbergstr.~1, 70569 Stuttgart, Germany}
\author{C.~Bernhard}
\affiliation{Max-Planck-Institute for Solid State Research,
Heisenbergstr.~1, 70569 Stuttgart, Germany}
\author{B.~Keimer}
\affiliation{Max-Planck-Institute for Solid State Research,
Heisenbergstr.~1, 70569 Stuttgart, Germany}
\author{L.~Alff}
\affiliation{Darmstadt
University of Technology, Petersenstr.~23, 64287 Darmstadt,
Germany}

\date{24th November 2005}

\begin{abstract}

The observation of superconductivity in the layered transition
metal oxide Na$_x$CoO$_2\cdot y$H$_2$O (K.~Takada, H.~Sakurai,
E.~Takayama-Muromachi, F.~Izumi, R.~A.~Dilanian, and T.~Sasaki,
Nature (London) {\bf 422}, 53 (2003)) has caused a tremendous
upsurge of scientific interest due to its similarities and its
differences to the copper based high-temperature superconductors.
Two years after the discovery, we report the fabrication of
single-phase superconducting epitaxial thin films of
Na$_{0.3}$CoO$_2\cdot 1.3D_2$O grown by pulsed laser deposition
technique. This opens additional roads for experimental research
exploring the superconducting state and the phase diagram of this
unconventional material.
\end{abstract}
\maketitle

Superconductivity in Na$_x$CoO$_2\cdot y$H$_2$O is a property of
the cobalt oxide planes. Like in the case of the cuprates, the
phase diagram of sodium cobaltate encompasses several competing
electronic phases \cite{Schaack:03,Milne:04,Ong:04}, and there are
many indications that the superconducting state is unconventional
\cite{Fujimoto:04,Bang:03}. While in high-temperature
superconductors the two-dimensional character of the corresponding
copper oxide planes is stabilized by the intrinsically strongly
anisotropic crystal structure, in Na$_x$CoO$_2$ a complicated
water intercalation process is needed to amplify this anisotropy
and induce superconductivity. The main difficulties in fabricating
high-quality bulk material of Na$_x$CoO$_2\cdot y$H$_2$O are the
following: First, sodium can be inhomogeneously distributed in the
entire bulk sample, resulting in an ill-defined doping level.
Second, impurity Co oxides such as CoO and Co$_3$O$_4$ are likely
to grow due to the high volatility of sodium. Third, it is
difficult to avoid the formation of Na$_2$CO$_3$ in conventional
bulk sample fabrication. As a result, high-quality single crystals
or bulk samples of Na$_x$CoO$_2\cdot y$H$_2$O are rare, and it is
even more difficult to obtain clean surfaces for optical
spectroscopy and tunnelling experiments in this material. While
high-quality thin films of high-temperature superconductors has
allowed a large variety of phase-sensitive experiments to explore
the symmetry of the superconducting order parameter, corresponding
studies have thus far not proven possible for water-intercalated
sodium cobaltate, due to the lack of superconducting thin films.
Thin-film deposition techniques offer major advantages in the
synthesis of high-quality samples, such as a well-defined vacuum
and oxidizing environment. Recently, epitaxial growth of
high-quality, "dry" Na$_x$CoO$_2$ films by pulsed laser deposition
on SrTiO3 substrates has been reported
\cite{Krockenberger:05,Ohta:05}. In these thin films, phase purity
was established within the detection limits of x-ray diffraction
in four-circle geometry, and flat surfaces are obtained. As a
further milestone, we now report the fabrication of
superconducting thin films of Na$_x$CoO$_2\cdot y$H$_2$O.

\begin{figure}[b]
\centering{%
\includegraphics[width=0.9\columnwidth,clip=]{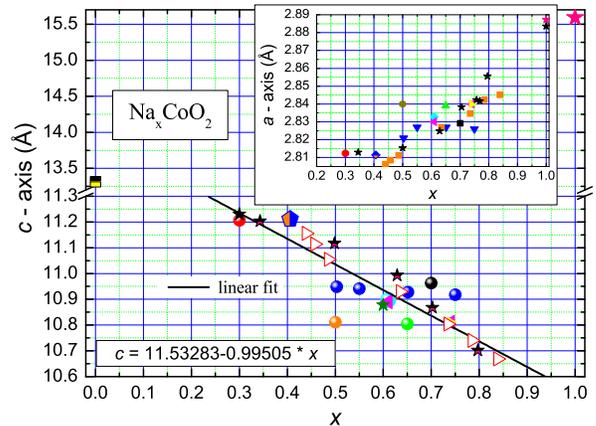}}
\caption{Lattice parameters from literature for independently
refined diffraction and ICP-AES measurements for different $x$ in
Na$_x$CoO$_2$. Between $x = 0.85$ and $0.95$ there is a symmetry
change of the crystal structure from point group 194 to 165. In
the doping range from $x = 0.25$ to $x = 0.85$ the $c$-axis
parameter varies linearly with doping. The $a$-axis parameter
varies only about 20\,pm throughout the whole doping
range.}\label{Fig:1}
\end{figure}

\begin{figure}[h]
\centering{%
\includegraphics[width=0.9\columnwidth,clip=]{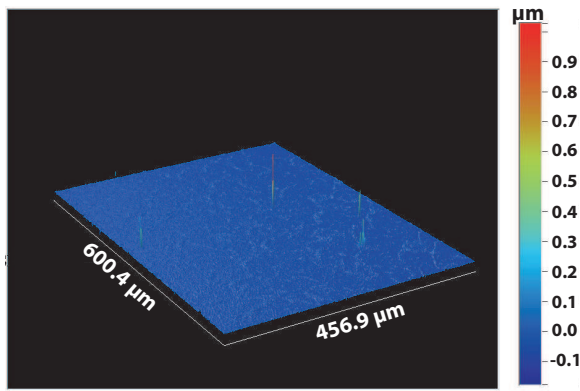}}
 \caption{White light interferometric scan image
of the surface of a Na$_{0.3}$CoO$_2\cdot1.3D_2$O thin film on
SrTiO$_3$ (001). Within the blue regions the surface roughness is
below 10\,nm. The rare spikes arise from sodium
carbonate.}\label{Fig:2}
\end{figure}

The control and the determination of the sodium contents are
essential to the growth of Na$_x$CoO$_2$. For bulk materials,
methods such as inductively coupled plasma - atomic emission
spectroscopy (ICP-AES) and energy dispersive x-ray (EDX) analysis
are available, but the results are often difficult to interpret
due to the presence of the above mentioned impurity phases. These
methods cannot be applied to thin films, because the material
quantity is too low. However, it is well known that for doped
oxide materials the lattice parameters depend strongly on the
doping level. A well-known example are the high-temperature
superconductors, see for instance Ref.~\cite{Naito:00}. For
Na$_x$CoO$_2$, the in-plane lattice constant $a$ shows almost no
dependence on doping, but doping does have a significant effect on
the c-axis parameter (see Fig.~\ref{Fig:1}). With increasing Na
contents, positive charge accumulates between the CoO$_2$ layers,
and as a result, the $c$-axis shrinks. The linear decrease of the
c-axis lattice parameter with increased doping is a useful and
accurate, albeit indirect, tool to determine the sodium contents
of Na$_x$CoO$_2$.

\begin{figure}[h]
\centering{%
\includegraphics[width=0.9\columnwidth,clip=]{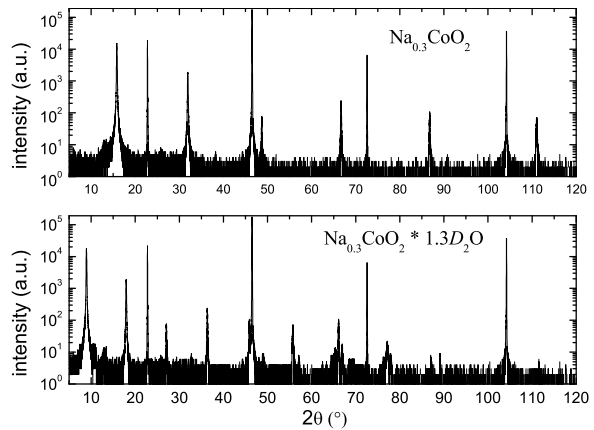}}
 \caption{X-ray diffraction pattern for epitaxial
Na$_{0.3}$CoO$_2\cdot1.3D_2$O on SrTiO$_3$ (001). Impurity phases
are below the detection limit.}\label{Fig:3}
\end{figure}

In order to obtain superconducting Na$_x$CoO$_2\cdot y$H$_2$O thin
films, it is necessary to intercalate water into samples with
doping levels of about $x = 0.3$. This goal can be achieved by
first stabilizing Na$_x$CoO$_2$ thin films with $x\simeq 0.6$ on
SrTiO$_3$ (001) substrates \cite{Krockenberger:05}. For obtaining
a film at the desired composition $x = 0.3$, it is vital to
provide a strong oxidizing agent to deintercalate Na$^+$ ions. In
principle, this can in be accomplished by the standard
Br$_2$-CH$_3$CN-solution method, which is also used for bulk
synthesis \cite{Takada:03}. However, we have found that
NO$_2$-BF$_4$ is superior to the conventional method, as Br is
avoided and the deintercalation time scale is considerably
accelerated. The decisive step is the intercalation of water into
the thin films. In contrast to bulk materials, single-phase thin
films cannot be simply immersed into water, because the thin films
tend to peel off the substrate. Only a much milder method allows
the successful fabrication of superconducting thin films. Oxygen
flow with 100\% humidity supplied by a $D_2$O bath at a
well-defined temperature (19$^\circ$C) is provided to the sample
over several days (196\,h). In contrast to alternative preparation
routes, this method also yields a clean and smooth surface (see
Fig.~\ref{Fig:2}), which is a necessary prerequisite for
meaningful surface-sensitive measurements. X-ray diffraction
clearly confirms the synthesis of correctly water-intercalated
sodium cobaltate Na$_x$CoO$_2\cdot yD_2$O epitaxial thin film with
the desired doping level $x = 0.3$ and water content $y = 1.3$
(see Fig.~\ref{Fig:3}).

\begin{figure}[b]
\centering{%
\includegraphics[width=0.9\columnwidth,clip=]{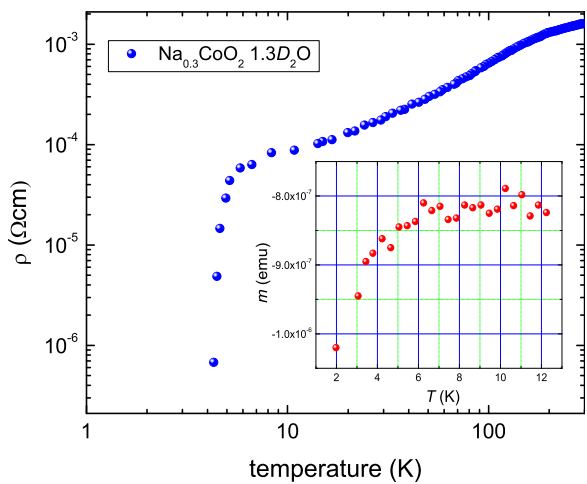}}
 \caption{Resistivity vs temperature for an epitaxial Na$_{0.3}$CoO$_2\cdot1.3D_2$O
 thin film on SrTiO$_3$ (001). In the inset superconducting quantum interference
 device (SQUID) magnetization measurement shows unambiguously the
flux expulsion effect at the critical temperature of about
$4.2$\,K.}\label{Fig:4}
\end{figure}

The water-intercalated Na$_x$CoO$_2\cdot yD_2$O  thin films indeed
show superconductivity for $x = 0.3$ and $y = 1.3$ with
$T_{\text{C,zero}}$ about $4.2$\,K (see Fig.~\ref{Fig:4}). The
relatively sharp superconducting transition with a width of
$1.5$\,K in the resistivity vs.~temperature curve confirms the
high quality of the thin films. Above the critical temperature,
the thin films show metallic behavior up to room temperature with
almost linear slope, similar to the high-temperature
superconductors.

\begin{figure}[h]
\centering{%
\includegraphics[width=0.9\columnwidth,clip=]{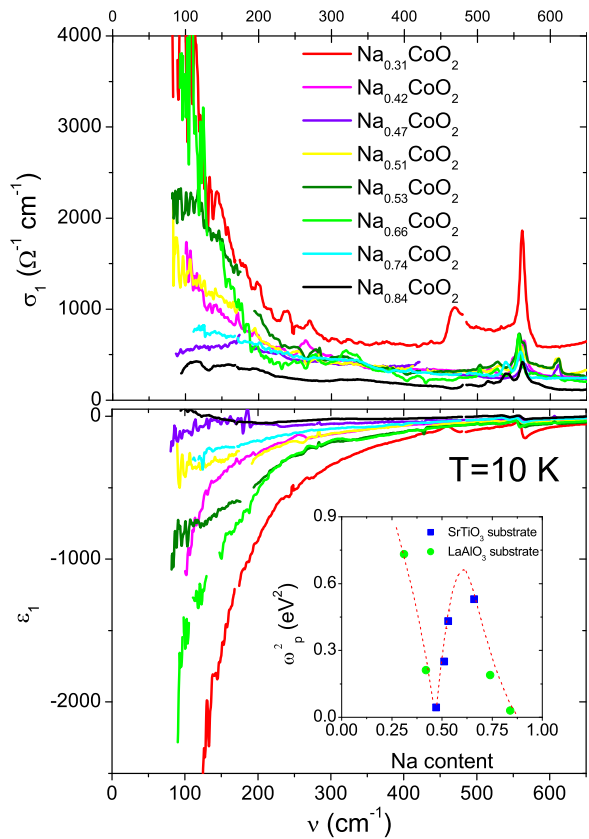}}
 \caption{Real parts of optical conductivity and dielectric function as a function of
 doping $x$ in Na$_x$CoO$_2$ films on SrTiO$_3$ (S) or LaAlO$_3$ (L) substrates.
 The free carrier plasma frequency, $\omega^2_{\text{p}}$ (see inset) clearly
 tracks the phase diagram of sodium cobaltate \cite{Ong:04} consisting of two
 metallic regions that are separated by an insulating region at
 half filling of $x = 1/2$.}\label{Fig:5}
\end{figure}

As a first flavor of the experimental possibilities which are
opened through the achievement of these high quality Na$_x$CoO$_2$
thin films we present in Fig.~\ref{Fig:5} the far-infrared
dielectric response function, $\varepsilon=\varepsilon_1+
i\varepsilon_2$, at $T = 10$\,K of a series of water-free samples
that cover a wide range of doping from $0.31 < x < 0.84$.
Displayed are the real parts of the optical conductivity,
$\sigma_1 = \omega/4\pi\varepsilon_2$ and of the dielectric
function, $\varepsilon_1$, as measured directly by spectral
ellipsometry using the ANKA synchrotron light source at
Forschungszentrum Karlsruhe \cite{Bernhard:04} (the contributions
of the SrTiO$_3$ or the LaAlO$_3$ substrates are subtracted). The
inset displays the evolution of the corresponding squared plasma
frequency, $\omega_{\text{p}}^2=\dfrac{4\pi n}{m^\ast}$, with $n$
the carrier density and $m^\ast$ their effective mass, that has
been deduced with a Drude-Lorentz fitting function. The free
carrier response evidently exhibits a very strong and
characteristic variation as a function of Na content. It has
fairly sharp maxima near $x = 1/3$ and $2/3$ that are separated by
a deep minimum near $x = 1/2$. It has indeed previously been noted
that the range of $x = 1/2$ separates two metallic regions with
high conductivity for $x < 0.5$ and $x > 0.5$
\cite{Schaack:03,Milne:04}. Notably, a correspondingly rapid
variation in the FIR electronic response was not observed in
previous optical measurements on single crystals
\cite{Lupi:05,Hwang:05}.  This suggests that the films can offer
significant advantages concerning the homogeneity (in particular
concerning the Na content) and the surface quality that is
critical for optical measurements. The high structural quality and
homogeneity of these films is also evident from very narrow phonon
modes in the FIR spectra whose analysis is omitted here.

The achievement of superconducting thin films of Na$_x$CoO$_2\cdot
yD_2$O paves the way for additional experiments with this
unconventional superconductor. Thin film based Josephson and
tunnelling experiments, which are expected to yield important
clues to the unconventional nature of superconductivity in this
material, can now readily be performed. Further, superconducting
quantum interference device (SQUID) experiments now come within
reach. Such experiments could confirm that, following the
discovery of $p$-wave superconductivity in Sr$_2$RuO$_4$
\cite{Nelson:04}, the compound Na$_x$CoO$_2\cdot y$H$_2$O is
another metallic solid-state analog to liquid $^3$He with an even
higher transition temperature. In a very recent paper, it was even
suggested that Na$_x$CoO$_2\cdot y$H$_2$O is an $f$-wave
superconductor \cite{Mazin:05} which would make this material the
most exotic one discovered so far. Following the successful
example of high-temperature superconductors, high-quality
superconducting thin film samples of water-intercalated sodium
cobaltate thus promise to enhance our knowledge about this complex
material.

\end{document}